\documentclass[aps,prd,showkeys,nofootinbib,floatfix,amsmath,amssymb,superscriptaddress,notitlepage,twocolumn]{revtex4-2}
\usepackage{rotating}
\usepackage{graphicx}
\usepackage{dcolumn}
\usepackage{bm}
\usepackage{color}
\usepackage{multirow}
\usepackage{mathtools}
\usepackage{siunitx}
\usepackage{slashed}

\usepackage[bookmarks=true]{hyperref}
\hypersetup{
   colorlinks,
   citecolor=black,
   filecolor=black,
   linkcolor=black,
   urlcolor=black
}
\usepackage{cleveref}

\graphicspath{{./}}

\begin{document}
\raggedbottom

\title{Electromagnetic form factors of singly charmed baryons from lattice QCD}
\author{Xing-yu Lan} %
\affiliation{Department of Physics, Hunan Normal University, Changsha 410081, China}
\author{Wen-zhen Hou} %
\affiliation{State Key Laboratory of Nuclear Physics and Technology, Institute of Quantum Matter, South China Normal University, Guangzhou 510006, China}
\affiliation{Key Laboratory of Atomic and Subatomic Structure and Quantum Control~(MOE), Guangdong-Hong Kong Joint Laboratory of Quantum Matter, Guangzhou 510006, China}
\affiliation{Guangdong Basic Research Center of Excellence for Structure and Fundamental Interactions of Matter, Guangdong Provincial Key Laboratory of Nuclear Science, Guangzhou 510006, China}
\author{Long-cheng Gui} %
\email{guilongcheng@hunnu.edu.cn}
\author{Wen Qin} %
\email{qinwen@hunnu.edu.cn}
\affiliation{Department of Physics, Hunan Normal University, Changsha 410081, China}
\affiliation{Hunan Research Center of the Basic Discipline for Quantum Effects and Quantum Technologies, Hunan Normal University, Changsha 410081, China}
\affiliation{Synergetic Innovation Center for Quantum Effects and Applications (SICQEA), Changsha 410081, China}
\affiliation{Key Laboratory of Low-Dimensional Quantum Structures and Quantum Control of Ministry of Education, Changsha 410081, China}
\author{Jian Liang} %
\email{jianliang@scnu.edu.cn}
\author{Jun Shi} %
\email{jun.shi@scnu.edu.cn}
\affiliation{State Key Laboratory of Nuclear Physics and Technology, Institute of Quantum Matter, South China Normal University, Guangzhou 510006, China}
\affiliation{Key Laboratory of Atomic and Subatomic Structure and Quantum Control~(MOE), Guangdong-Hong Kong Joint Laboratory of Quantum Matter, Guangzhou 510006, China}
\affiliation{Guangdong Basic Research Center of Excellence for Structure and Fundamental Interactions of Matter, Guangdong Provincial Key Laboratory of Nuclear Science, Guangzhou 510006, China}
\author{Long-Sheng Lu}
\email{lulongsheng@tsu.edu.cn}
\affiliation{Department of Physics, Taishan University, Taian 271000, China}
\begin{abstract}
We study the electromagnetic structure of the singly charmed baryons $\Sigma_{c}^{0},\Sigma_{c}^{+},\Sigma_{c}^{++}$ and $\Omega_{c}^{0}$ using six $(2+1)$-flavor gauge ensembles with Wilson--clover fermions at three lattice spacings and pion masses extending down to near the physical value. Joint fits to two-point and three-point correlation functions yield the flavor-separated electric and magnetic form factors. Their $Q^2$ dependence is described by the $z$ expansion to extract mean-square charge and magnetic radii and magnetic moments on each ensemble. Combined continuum and chiral fits then determine these observables at the physical pion mass in the continuum limit. The charm-quark contribution to the magnetic moment is substantially smaller in magnitude than the light- and strange-quark contributions, with $|\mu_{c}/\mu_{l}|=0.082(24)$ for $\Sigma_{c}$ and $|\mu_{c}/\mu_{s}|=0.097(16)$ for $\Omega_{c}^{0}$, consistent with heavy-quark expectations. The light- and strange-quark charge distributions are more spatially extended than the charm-quark distribution. Our results also show appreciable finite-lattice-spacing effects in the electromagnetic radii.

\end{abstract}

\maketitle
\section{Introduction}
Charmed baryons combine a heavy quark with light degrees of freedom, providing a setting for studying the nonperturbative dynamics of quantum chromodynamics (QCD).
Recent experimental progress on the electromagnetic properties of charmed baryons provides important input for theoretical studies.
BESIII measurements of the $\Lambda_c^+\bar{\Lambda}_c^-$ production cross section, angular distributions, and transverse polarization provide information on the magnitudes and relative phase of the timelike electric and magnetic form factors~\cite{BESIII:2023rwv,BESIII:2026LcPolarization}.
For $\Sigma_c$, BESIII performed the first search for $e^+e^-\to\Sigma_c\bar{\Sigma}_c$ near threshold and set upper limits on the Born cross section~\cite{BESIII:2025srf}, providing experimental constraints on its timelike electromagnetic structure.
In the charmed-strange sector, BaBar and Belle studied radiative transitions between the excited and ground states through $\Omega_c^{*0}\to\Omega_c^0\gamma$~\cite{BaBar:2006pve,Solovieva:2008fw}.

Baryon electromagnetic form factors encode information on charge and magnetization distributions and quark dynamics,
making them important probes of baryon structure.
Quark models, QCD sum rules, and chiral effective theories have been widely used to study
the magnetic moments and electromagnetic form factors of singly charmed baryons%
~\cite{Mohan:2022sxm,Wang:2018gpl,Shi:2018rhk,Aliev:2015axa,
Yang:2018uoj,Faessler:2006ft,Zhu:1997as,Kucukyilmaz:2025rsd,
Sharma:2010vv,Cao:2026EMFF}.
However, quantitative predictions for their magnetic moments and charge radii differ substantially among theoretical approaches.
For example, the bag-model predictions for the $\Sigma_c^{++}$ and $\Omega_c^0$ magnetic moments are
$1.68\,\mu_N$ and $-0.77\,\mu_N$, respectively~\cite{Bernotas:2012nz},
whereas the corresponding chiral constituent quark model results are
$2.20\,\mu_N$ and $-0.90\,\mu_N$~\cite{Sharma:2010vv}.
The chiral quark-soliton model predicts a mean-square charge radius of $0.59\,\mathrm{fm}^2$ for $\Sigma_c^{++}$~\cite{Kim:2018nqf}, compared with $0.192(22)\,\mathrm{fm}^2$ from an earlier lattice calculation~\cite{Can:2013tna}.
First-principles lattice QCD calculations of the electromagnetic properties of singly charmed baryons are therefore needed
to test these theoretical descriptions.

Lattice studies of charmed baryons have yielded results for the mass spectrum~\citep{Hu:2024mas,Chiu:2005zc,PACS-CS:2013vie,Brown:2014ena,Briceno:2012wt,Alexandrou:2023dlu}, mass decomposition~\citep{Li:2022vbc}, and semileptonic form factors and the associated CKM matrix elements~\citep{FlavourLatticeAveragingGroupFLAG:2024oxs,Zhang:2021oja}, whereas calculations of electromagnetic form factors remain comparatively scarce. Can et al. computed elastic electromagnetic form factors of baryons including $\Sigma_c$ and $\Omega_c$ using $(2+1)$-flavor gauge ensembles of size $32^3\times64$ at a single lattice spacing, $a\simeq0.096\,\mathrm{fm}$, and pion masses $m_\pi\simeq300$, $410$, $570$, and $700\,\mathrm{MeV}$. They extracted mean-square charge and magnetic radii and magnetic moments~\cite{Can:2013tna}. A subsequent study investigated the electromagnetic properties of $\Omega_c$ and related charmed-strange baryons on a $32^3\times64$ ensemble with $a=0.0907(13)\,\mathrm{fm}$ and $m_\pi\simeq156\,\mathrm{MeV}$~\cite{Can:2015exa}. These calculations resolved the quark-sector contributions and revealed the distinct roles of light, strange, and charm quarks in baryon charge radii and magnetic moments. Related lattice studies also cover elastic form factors of doubly charmed baryons and radiative transitions of charmed baryons~\cite{Can:2013zpa,Bahtiyar:2015sga,Bahtiyar:2016dom,
Bahtiyar:2018vub,Bahtiyar:2022nqw,Can:2021ehb}.

In this work, we study the spacelike electromagnetic form factors of $\Sigma_c^0$, $\Sigma_c^+$, $\Sigma_c^{++}$, and $\Omega_c^0$ on six $(2+1)$-flavor gauge ensembles with Wilson--clover fermions. The three lattice spacings span $0.05187(26)$--$0.10530(18)\,\mathrm{fm}$, and the pion masses extend down to near the physical value. We extract the flavor-separated electric and magnetic form factors from joint fits to two-point and three-point correlation functions. The $z$ expansion determines their values and slopes at zero momentum transfer, from which we obtain radii and magnetic moments on each ensemble. These observables are then extrapolated to the continuum limit and the physical pion mass.

Section~\ref{sec:theo} presents the theoretical formalism, Sec.~\ref{sec:result} describes the numerical calculation, Sec.~\ref{sec:discussion} discusses the results, and Sec.~\ref{sec:conclusions} summarizes our findings.

\section{Theoretical formalism}
\label{sec:theo}
\subsection{Matrix elements}
For a baryon $\mathcal B$, the matrix element of each flavor vector current is parameterized as
\begin{align}
 \langle\mathcal B(p_f,s_f)|&J_\mu^f|
 \mathcal B(p_i,s_i)\rangle\nonumber\\
 &=\bar u(p_f,s_f)\,\mathcal O_\mu^f(Q^2)\,u(p_i,s_i),
 \label{eq:matrix_element_definition}
\end{align}
\begin{equation}
 \mathcal O_\mu^f(Q^2)=\gamma_\mu F_{1,\mathcal B}^f(Q^2)
 +i\frac{\sigma_{\mu\nu}q^\nu}{2m_{\mathcal B}}F_{2,\mathcal B}^f(Q^2).
 \label{eq:vertex_definition}
\end{equation}
Here $q=p_f-p_i$ and $Q^2=-q^2$. The vector current for flavor $f$ with unit electric charge is defined by
\begin{equation}
 J_\mu^f=\bar q_f\gamma_\mu q_f.
 \label{eq:veccurrrent}
\end{equation}
The corresponding Sachs form factors are
\begin{align}
G^f_{E,\mathcal B}(Q^2)&=F^f_{1,\mathcal B}(Q^2)
 -\frac{Q^2}{4m_{\mathcal B}^2}F^f_{2,\mathcal B}(Q^2),\\
G^f_{M,\mathcal B}(Q^2)&=F^f_{1,\mathcal B}(Q^2)+F^f_{2,\mathcal B}(Q^2).
\end{align}

\subsection{Mean-square radii and magnetic moments}
The flavor-separated mean-square charge and magnetic radii are determined by the slopes of the normalized form factors at $Q^2=0$,
\begin{equation}
    \left\langle r^{2,f}_{E,M}\right\rangle
    =
    -\frac{6}{G^f_{E,M}(0)}
    \frac{dG^f_{E,M}(Q^2)}{dQ^2}\bigg|_{Q^2=0}.
\label{eq:charge_radii_z}
\end{equation}

The physical electromagnetic form factors are obtained by summing the flavor contributions weighted by the quark electric charges:
\begin{equation}
G_{E,M}(Q^2)
=
\sum_f e_f G^f_{E,M}(Q^2),
\label{eq:form_factor_add}
\end{equation}
where $e_f$ is the quark charge in units of $e$, and the sum runs over $u,d,s,c$.
The mean-square charge radius of a charged baryon is
\begin{equation}
\left\langle r_E^2\right\rangle
=-\frac{6}{G_E(0)}
\frac{dG_E(Q^2)}{dQ^2}\bigg|_{Q^2=0}.
\label{eq:charge_radius}
\end{equation}
For a neutral baryon, the mean-square charge radius is defined directly from the slope of the total electric form factor:
\begin{equation}
\left\langle r_E^2\right\rangle
=
-6\frac{dG_E(Q^2)}{dQ^2}\bigg|_{Q^2=0}.
\label{eq:neutral_charge_radius}
\end{equation}
The magnetic moment associated with flavor $f$ is obtained from its magnetic form factor at zero momentum transfer,
\begin{equation}
\mu_{\mathcal{B}}^f
=
G_M^f(0)
\frac{m_N}{m_{\mathcal{B}}}\mu_N ,
\label{eq:mu}
\end{equation}
where $\mu_N$ is the nuclear magneton, $m_N$ is the physical nucleon mass, and $m_{\mathcal{B}}$ is the baryon mass determined on the lattice. The factor $m_N/m_{\mathcal{B}}$ converts the magnetic moment to nuclear magnetons.
The physical baryon magnetic moment is obtained from the charge-weighted sum of the flavor contributions:
\begin{equation}
\mu_{\mathcal{B}}
=
\frac{m_N}{m_{\mathcal{B}}}
\sum_f e_f G_M^f(0)\mu_N .
\label{eq:total_mu}
\end{equation}

\subsection{Correlation functions}
The baryon matrix elements defined in Eq.~\eqref{eq:matrix_element_definition} are extracted from Euclidean two-point and three-point correlation functions on the lattice:
\begin{equation}
\label{eq:twopt}
C^{(2)}(\vec{p},t)=\sum_{\vec{x}}e^{-i\vec{x}\cdot\vec p}\operatorname{Tr}\left[\Gamma_4\langle\chi(t,\vec{x})\bar{\chi}(0)\rangle\right],
\end{equation}
\begin{multline}
C^{(3)}_{\mu}(\Gamma_{\nu};\vec{p}_f,\vec{q},\vec{p}_i;0,t,t_f)
=\sum_{\vec{x},\vec{x}_f} e^{-i\vec{x}_f\cdot\vec{p}_f}
 e^{i\vec{q}\cdot\vec{x}}\\
\times\operatorname{Tr}\!\left[
\Gamma_\nu\langle\chi(t_f,\vec{x}_f)j^f_\mu(t,\vec{x})
\bar\chi(0,\vec{0})\rangle\right].
\label{eq:3pt_hadron}
\end{multline}
In Eq.~\eqref{eq:3pt_hadron}, $\Gamma_\nu$ denotes the spin-parity projector, with $\Gamma_4=\frac{1+\gamma_4}{2}$ and $\Gamma_i=i\Gamma_4\gamma_5\gamma_i$. The trace acts on Dirac indices, $j_\mu^f=\bar q_f\gamma_\mu q_f$ is the vector current for flavor $f$, and $\mu$ and $\nu$ are Lorentz indices. The Dirac adjoint of the interpolating operator is defined as $\bar\chi=\chi^\dagger\gamma_4$. We use the following interpolating operator for singly charmed baryons:
\begin{eqnarray}
\chi=\epsilon_{abc}(q^T_{a}(C\gamma_5)c_b)q_{c},
\end{eqnarray}
where $q$ denotes a light- or strange-quark field, depending on the baryon, $a$, $b$, and $c$ are color indices, and $C=\gamma_4\gamma_2$ is the charge-conjugation matrix.

The spectral decompositions of the two-point and three-point correlation functions are
\begin{align}
 C^{(2)}(\vec p,t)
 &=\sum_{n,s} e^{-E_n t}\operatorname{Tr}\!\left[
 \Gamma_4\langle\Omega|\chi(\vec p)|n,s\rangle\right.\nonumber\\
 &\qquad\left.\times\langle n,s|\bar\chi(\vec p)|\Omega\rangle\right],
 \label{eq:2pt_hadron_real}
\end{align}
\begin{align}
 C_\mu^{(3)}(\Gamma_\nu,\vec q,t,t_f)
 &=\sum_{m,n,s',s}e^{-E_m(t_f-t)}e^{-E_n t}\nonumber\\
 &\quad\times\operatorname{Tr}\!\left[
 \Gamma_\nu\langle\Omega|\chi(\vec p_f)|m,s'\rangle\right.\nonumber\\
 &\qquad\times\langle m,s'|j_\mu^f(\vec q)|n,s\rangle\nonumber\\
 &\qquad\left.\times\langle n,s|\bar\chi(\vec p_i)|\Omega\rangle\right].
 \label{eq:3pt_hadron_real}
\end{align}
We absorb the state and spinor normalization factors into the overlap amplitudes and extract the ground-state matrix elements from joint fits to the two-point and three-point correlation functions. We denote the spin-projected, renormalized ground-state matrix element by $\Pi_\mu^f(\vec p_f,\vec p_i;\Gamma_\nu)$. With the final baryon at rest, $\vec p_f=0$, $E_f=m_{\mathcal B}$, and $\vec p_i=-\vec q$, its relation to the Sachs form factors is

  \begin{align}
  \Pi_4^f(0,-\vec{q},\Gamma_4)&=\left[\frac{(E_i + E_f)}{2E_i}\right]^{\frac{1}{2}}G_{E,\mathcal{B}}^f(Q^2),
  \label{eq:RE}\\
     \Pi_i^f(0,-\vec{q},\Gamma_j)&=\left[\frac{1}{2E_i (E_i+ E_f)}\right]^{\frac{1}{2}}\epsilon_{ijk}q_kG_{M,\mathcal{B}}^f(Q^2).
     \label{eq:RM}
  \end{align}

\section{Numerical details}
\label{sec:result}
We use six $(2+1)$-flavor gauge ensembles with Wilson--clover fermions generated by the CLQCD Collaboration~\cite{CLQCD:2023sdb}, with parameters listed in Table~\ref{tab:ensemble}. The coarse, intermediate, and fine lattice spacings are approximately $0.105$, $0.077$, and $0.052\,\mathrm{fm}$, respectively, and the pion masses range from $135.5$ to $317.2\,\mathrm{MeV}$. Varying both the lattice spacing and the light-quark mass allows us to study discretization effects and quark-mass dependence in the flavor-separated matrix elements. C48P14 also provides the smallest nonzero momentum transfer, helping to constrain the form-factor slopes near $Q^2=0$.

%%%%%%%%%%%%%%%%%%%%%%%%%%%%%%%%%%%%%%%
\begin{table*}[!htbp]
\centering
\caption{Parameters of the gauge ensembles used in this work. From top to bottom, the rows give the ensemble names, lattice spacings $a$, bare quark masses (light $a\mu_l$, strange $a\mu_s$, and charm $a\mu_c$), lattice sizes $L^3\times T$, pion masses $m_\pi$, and numbers of configurations $N_{\mathrm{cfg}}$.}
\label{tab:ensemble}
\renewcommand{\arraystretch}{1.4}
\small
\begin{tabular*}{\textwidth}{@{\extracolsep{\fill}}lcccccc@{}}  %
\hline \hline
\textbf{Ensemble}\quad & \textbf{C32P23} & \textbf{C24P29} & \textbf{C48P14}& \textbf{F32P21} & \textbf{F32P30} & \textbf{H48P32} \\
\hline
$a$~\text{(fm)} & 0.10530(18) & 0.10530(18) & 0.10530(18)& 0.07746(18) & 0.07746(18) & 0.05187(26) \\
$a\mu_l$ & -0.2790 & -0.2770 & -0.2825 & -0.2320 & -0.2295 & -0.1850 \\
$a\mu_s$ & -0.2400 & -0.2400 & -0.2310& -0.2050 & -0.2050 & -0.1700 \\
$a\mu_c$ & 0.4500 & 0.4479 & 0.4502& 0.2104 & 0.2079 & 0.0581 \\
$L^3 \times T$ & $32^3\times64$ & $24^3\times72$ & $48^3 \times 96$ & $32^3\times64$ & $32^3\times96$ & $48^3 \times 144$ \\
$m_{\pi}$~\text{(MeV)} & 228.0(1.2) & 292.7(1.2) &135.5(1.6)& 210.9(2.2) & 303.4(0.9) & 317.2(0.9) \\
%$m_{\pi}L$ & 3.894(20) & 3.746(13) &3.495(44) & 2.646(28) &3.780(15) &4.000(12) \\
$N_{\mathrm{cfg}}$ & $400$ & $450$ & $500$ & $400$ & $302$ & $200$\\
\hline \hline
\end{tabular*}
\end{table*}
%%%%%%%%%%%%%%%%%%%%%%%%%%%%%%%%%%%%%%%%%%%%%%
\subsection{Two-point correlation functions}
\label{Two-point correlation functions}
%%%%%%%%%%%%%%%%%%%%%%%%%%%%%%

\begin{figure*}[!tp]
\centering
\includegraphics[width=\textwidth]{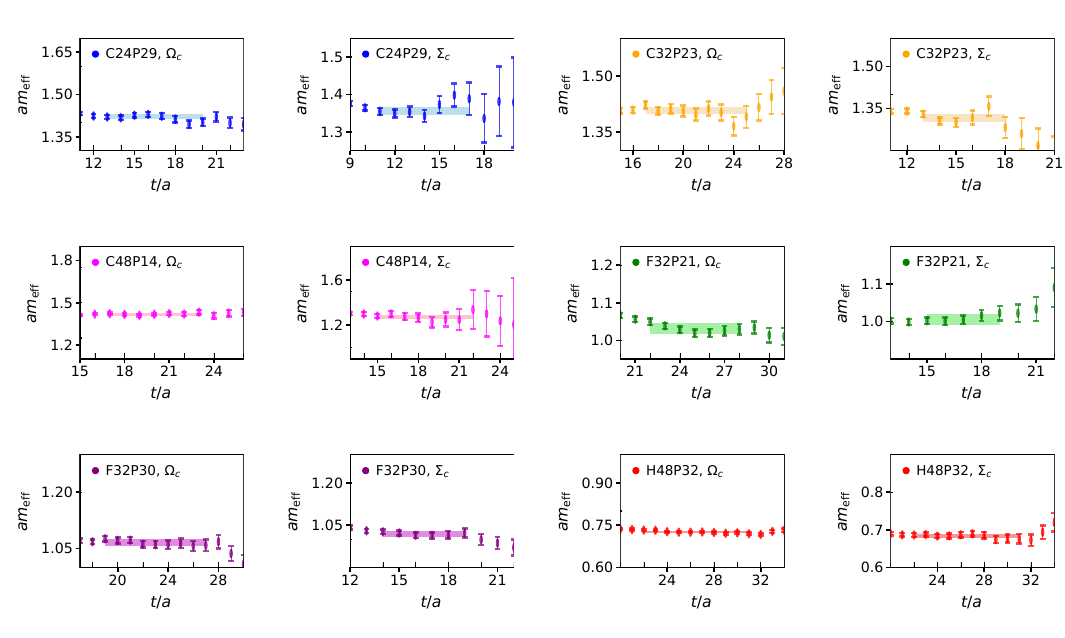}
\caption{Effective masses of the $\Omega_c$ and $\Sigma_c$ baryons on the six gauge ensembles. Each adjacent pair shows $\Omega_c$ on the left and $\Sigma_c$ on the right for the ensemble identified in the data legend. The bands show the fitted ground-state masses and their uncertainties.}
\label{fig:baryon_masses}
\end{figure*}

We use Gaussian-smeared quark sources to improve the overlap of the baryon interpolating operators with the ground state. The two-point functions used in the form-factor analysis employ the same smearing at the source and sink. We use the charm-quark mass parameters adopted in Ref.~\cite{Hou:2026xif}, which were tuned to reproduce the physical $J/\psi$ mass~\cite{Meng:2024gpd}.

%For the consistency of subsequent three-point function analyses, we apply the same smearing procedure to the fermion sinks.

To extract the ground-state mass, we define the effective mass as
\begin{equation}
    aE_{\text{eff}}(t) = \ln \frac{C_2(t)}{C_2(t+1)}
    \label{eq:eff_mass_definition}
\end{equation}
We select fit intervals from the effective-mass plateaus and fit the two-point correlation functions with a single exponential to extract the ground-state masses. Figure~\ref{fig:baryon_masses} shows the effective masses and fit results on all six ensembles. The horizontal extent of each band indicates the fit interval, and its vertical extent shows the statistical uncertainty. Table~\ref{tab:mass} summarizes the extracted masses, experimental values, and results from other lattice collaborations.

%The single-exponential fits here determine the masses; the form factors below are extracted from joint two-state fits to the two-point and three-point correlation functions.

%%%%%%%%%%%%%%%%%%%%%%%%%%%%%%%%
\subsection{Three-point correlation functions}

In the Wick contractions of the three-point functions, we compute only the connected-insertion correlation functions, in which the electromagnetic current couples to a valence-quark line in the baryon. We use the sequential-source method through the sink~\cite{LHPC:2002xzk}, fixing the sink momentum, spin projection, and source-sink separation. The sink is at rest and the source time is $t_i=0$; the magnetic correlation functions use the spin projection $\nu=3$. The spatial momenta are $\vec p=(2\pi/L)\vec n$, with $|\vec n|^2\leq6$. Equivalent momentum directions are combined with the signs of the magnetic projections in Eq.~\eqref{eq:RM}, retaining correlations within each common data source.
The squared four-momentum transfer is $Q^2=|\vec p_i|^2-(E_{\rm kin}-m_{\mathcal B})^2=2m_{\mathcal B}(E_{\rm kin}-m_{\mathcal B}).$ Here $E_{\rm kin}=\sqrt{m_{\mathcal B}^2+|\vec p_i|^2}$.

The source-sink separations are $t_f/a=10,12,14$ on the coarse ensembles, $12,14,16$ on the intermediate ensembles, and $14,16,18$ on H48P32. Data at multiple source-sink separations improve control of excited-state contamination.

We perform joint two-state fits to the two-point and three-point correlation functions. The superscript $\mathrm{ss}$ denotes smearing at both the source and sink of the two-point functions. Absorbing the kinematic normalization factors associated with the spin projection into the overlap amplitudes gives
\begin{align}
    C^{(2),\mathrm{ss}}(\vec{p}_i,t) &\approx |A_0|^2 e^{-E_i t} + |A_1|^2 e^{-E^\prime_i t},
    \label{eq:2pt_joint_source}\\
    C^{(2),\mathrm{ss}}(\vec{p}_f,t) &\approx |B_0|^2 e^{-E_f t} + |B_1|^2 e^{-E^\prime_f t},
    \label{eq:2pt_joint_sink}
\end{align}

\begin{eqnarray}
    C^{(3)}_\mu(& &\Gamma_{\nu},\vec{q},t_f,t) \approx  A_0\Pi_{\mu,\mathrm{bare}}^f B_0^* e^{-E_f(t_f-t)} e^{-E_it} \nonumber \\
    +& &A_0\Pi_{\mu,\mathrm{bare}}^{f,\prime} B_1^* e^{-E_f^\prime(t_f-t)} e^{-E_it} \nonumber \\
    +& &A_1 \Pi_{\mu,\mathrm{bare}}^{f,\prime \prime} B_0^* e^{-E_f(t_f-t)} e^{-E_i^\prime t} \nonumber \\
    +& &A_1 \Pi_{\mu,\mathrm{bare}}^{f,\prime \prime \prime} B_1^* e^{-E_f^\prime (t_f-t)} e^{-E_i^\prime t}. \label{3pt_joint}
\end{eqnarray}
The parameters $E_i$ and $E_f$ denote the ground-state energies in the initial and final channels, while $E_i'$ and $E_f'$ are effective energy parameters that account for excited-state contributions in the respective channels. We fit the ground-state energies and effective energy gaps independently in the two channels. The overlap amplitudes $A_n$ and $B_n$ include normalization factors and the couplings of the interpolating operators to the initial and final states; subscripts $0$ and $1$ label the ground-state and effective excited-state terms. In this parameterization of the bare correlation functions, $\Pi_{\mu,\mathrm{bare}}^f$ is the bare ground-state matrix element, while $\Pi_{\mu,\mathrm{bare}}^{f,\prime}$, $\Pi_{\mu,\mathrm{bare}}^{f,\prime\prime}$, and $\Pi_{\mu,\mathrm{bare}}^{f,\prime\prime\prime}$ are transition matrix elements involving excited states, each treated as an independent fit parameter. The subscript $\mathrm{bare}$ denotes a matrix element before multiplication by the current renormalization constant $Z_V$.

\begin{figure}[!htbp]
\centering
\includegraphics[width=\linewidth]{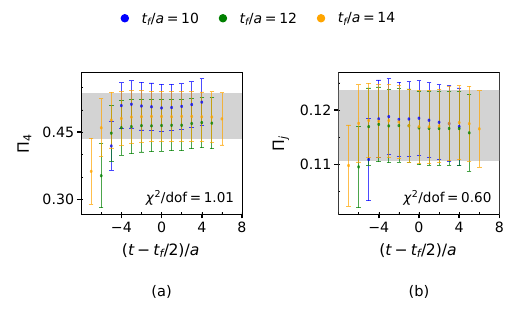}
\caption{Bare electric (left) and magnetic (right) matrix elements extracted from joint fits on a coarse ensemble, with source-sink separations $t_f/a=10,12,14$. The points are obtained after subtracting the fitted excited-state contributions, and the gray bands show the fitted bare ground-state matrix elements. The labels $\Pi_4$ and $\Pi_j$ denote $\Pi_{4,\mathrm{bare}}^f$ and $\Pi_{j,\mathrm{bare}}^f$, respectively. Colors distinguish the source-sink separations. The horizontal coordinate is measured relative to the source-sink midpoint.}
\label{fig:bare_fits}
\end{figure}

Figure~\ref{fig:bare_fits} shows the bare electric and magnetic matrix elements extracted from the joint fits. We vary the temporal fit windows and include the resulting systematic uncertainty in the total uncertainty of the matrix elements.

The flavor vector current is conserved in the continuum theory. At nonzero lattice spacing,
the local vector current used here is not exactly conserved and requires a renormalization factor
$Z_V$ to satisfy the corresponding charge normalization condition:
\[
J_\mu^{f,R}=Z_V j_\mu^f.
\]
We use the same vector-current renormalization constant $Z_V$ as Hou et al.~\cite{Hou:2026xif}.
The renormalized matrix element is $\Pi_\mu^f=Z_V\Pi_{\mu,\mathrm{bare}}^f$.

\section{Results and discussion}
\label{sec:discussion}
\subsection{Mass spectrum}

\begin{figure*}[!tp]
\centering
\includegraphics[width=\textwidth]{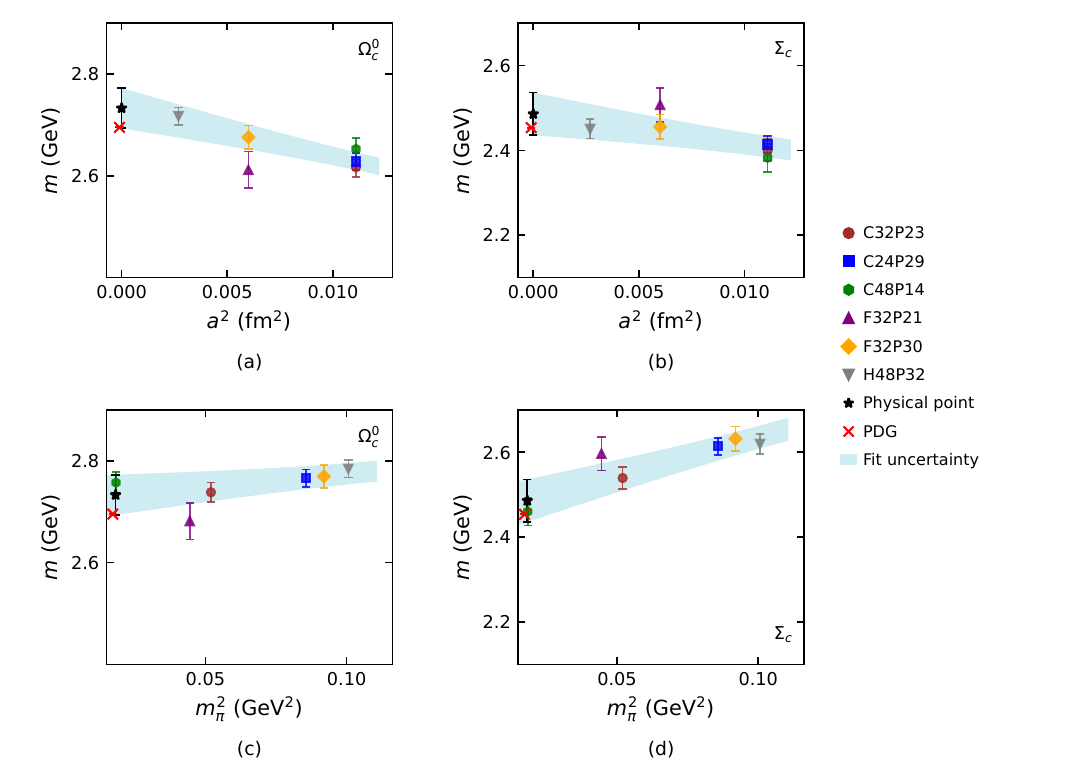}
\caption{Combined continuum and chiral extrapolations of the $\Omega_c$ (left column) and $\Sigma_c$ (right column) ground-state masses. The upper panels show the $a^2$ dependence at the physical pion mass, and the lower panels show the $m_\pi^2$ dependence in the continuum limit. The shared legend identifies the ensembles by symbol and color. Black stars denote the results in the continuum limit at the physical pion mass, and bands show the fit uncertainties. Red crosses denote the PDG values and are shifted slightly to the left for visibility. For each projection, the points are shifted by subtracting the fitted contribution of the other variable.}
\label{fig:eff-wt}
\end{figure*}

We perform a combined continuum and chiral extrapolation of the $\Omega_c$ and $\Sigma_c$ ground-state masses extracted on the six ensembles, using the fit form
\begin{multline}
 f(a^2,m_\pi^2)=\mathcal M_B(0,m_{\pi,\mathrm{phy}}^2)+c_1a^2\\
 +c_2(m_\pi^2-m_{\pi,\mathrm{phy}}^2).
 \label{eq:mass_wt}
\end{multline}
Here $\mathcal M_B(0,m_{\pi,\mathrm{phy}}^2)$ is the baryon mass in the continuum limit at the physical pion mass, $m_{\pi,\mathrm{phy}}=0.135\,\mathrm{GeV}$, and the fit parameters $c_1$ and $c_2$ describe the $a^2$ and $m_\pi^2$ dependence, respectively. Figure~\ref{fig:eff-wt} shows the combined extrapolation for $\Omega_c$ in the left column and $\Sigma_c$ in the right column. The upper panels show the $a^2$ dependence at the physical pion mass, and the lower panels show the $m_\pi^2$ dependence in the continuum limit. The results in the continuum limit at the physical pion mass are $m_{\Omega_c}=2.733(40)\,\mathrm{GeV}$ and $m_{\Sigma_c}=2.486(50)\,\mathrm{GeV}$, consistent within uncertainties with the experimental values compiled by the PDG~\cite{ParticleDataGroup:2024cfk}. Table~\ref{tab:mass} lists the ensemble results, the extrapolated values, and other lattice QCD determinations.

\subsection{Electric properties}
We parameterize the $Q^2$ dependence of the electromagnetic form factors using a truncated $z$ expansion~\cite{Hill:2010z}:
\begin{equation}
G_{E,M}^{f}(Q^2) = \sum_{k=0}^{k_{\rm max}} a_k^{f} z(Q^2)^k ,
\label{eq:z_expansion}
\end{equation}
where the conformal variable $z$ is defined as
\begin{equation}
z(Q^2) = \frac{\sqrt{t_{\rm cut}+Q^2}-\sqrt{t_{\rm cut}}} {\sqrt{t_{\rm cut}+Q^2}+\sqrt{t_{\rm cut}}},
\label{eq:z_definition}
\end{equation}
We set $t_{\rm cut}=4m_D^2$, where $m_D$ is the $D$-meson mass, and fit the flavor-separated electric and magnetic form factors using a cubic $z$ expansion ($k_{\rm max}=3$). The fitted form factors and their derivatives at $Q^2=0$ determine the mean-square charge and magnetic radii through Eq.~\eqref{eq:charge_radii_z}.

\begin{figure*}[!tp]
\begin{minipage}[t]{\columnwidth}
\centering
%\subfigure{
\includegraphics[width=\linewidth]{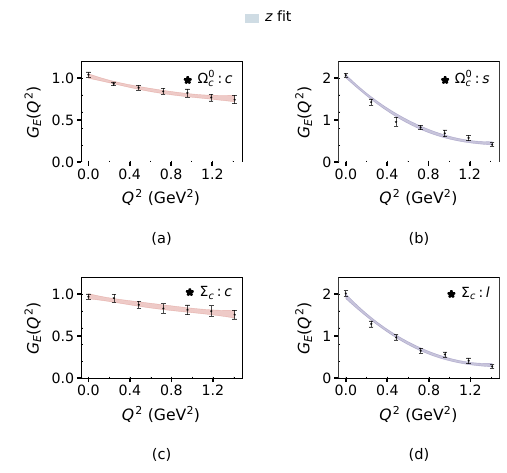}
%}
\caption{$z$-expansion fits to the flavor-separated electric form factors $G_E^f(Q^2)$ on H48P32. The flavor-separated form factors include all valence quarks of the indicated flavor, with the electric charge factor removed. From left to right, the upper panels show the charm- and strange-quark contributions in $\Omega_c$, and the lower panels show the charm- and light-quark contributions in $\Sigma_c$. Points are lattice results, and bands show the $z$-expansion fits and their uncertainties. The labels $c,s,l$ in the data legends identify the charm, strange, and light sectors, respectively.}
\label{fig:quark_E}
\end{minipage}\hfill
\begin{minipage}[t]{\columnwidth}
\centering
%\subfigure{
\includegraphics[width=\linewidth]{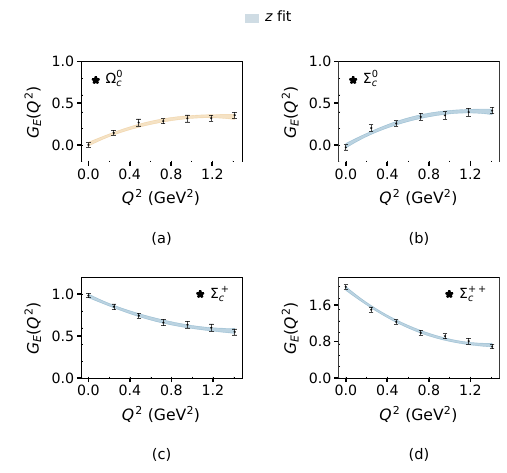}
%}
\caption{$z$-expansion fits to the total electric form factors on H48P32, obtained from the charge-weighted sums of the flavor contributions. From left to right, the upper panels show $\Omega_c^0$ and $\Sigma_c^0$, and the lower panels show $\Sigma_c^+$ and $\Sigma_c^{++}$. Points are lattice results, and bands show the $z$-expansion fits and their uncertainties.}
\label{fig:whole_E}
\end{minipage}
\end{figure*}

Figure~\ref{fig:quark_E} shows the flavor-separated electric form factors and their $z$-expansion fits on H48P32. The light- and strange-quark electric form factors decrease more rapidly with $Q^2$ than the charm-quark form factors. The corresponding flavor-separated mean-square radii are listed in Table~\ref{tab:electric_radii}.

We denote the total contributions of the two light valence quarks in $\Sigma_c$ to the electric and magnetic form factors by $G_E^{l,\Sigma_c}$ and $G_M^{l,\Sigma_c}$, respectively, and the corresponding magnetic moment by $\mu_l$. Combining the flavor contributions with the quark electric charges gives the total baryon electric form factors shown in Fig.~\ref{fig:whole_E}. The mean-square charge radii of charged and neutral baryons are obtained from Eqs.~\eqref{eq:charge_radius} and~\eqref{eq:neutral_charge_radius}, respectively.

%%%%%%%%%%%%%%%%%%%%%%%%%%%%%%%%%%%%%%%%%

%%%%%%%%%%%%%%%%%%%%%%%%%%%%%%%%%%%%%%%%%%%%%%%%

Table~\ref{tab:electric_radii} quantifies the differences between the quark charge distributions. On H48P32, for example, the strange- and charm-quark mean-square charge radii in $\Omega_c$ are $0.268(16)$ and $0.079(23)\,\mathrm{fm}^2$, respectively, while the corresponding light- and charm-quark values in $\Sigma_c$ are $0.292(16)$ and $0.052(30)\,\mathrm{fm}^2$. These results show that the charm-quark charge distribution is more localized than the light- and strange-quark distributions. The $\Omega_c^0$ and $\Sigma_c^0$ mean-square charge radii are negative on all ensembles, reflecting the more extended charge distributions of the negatively charged strange or light quarks relative to that of the positively charged charm quark.

\subsection{Magnetic properties}
\begin{figure*}[!tp]
\begin{minipage}[t]{\columnwidth}
\centering
%\subfigure{
\includegraphics[width=\linewidth]{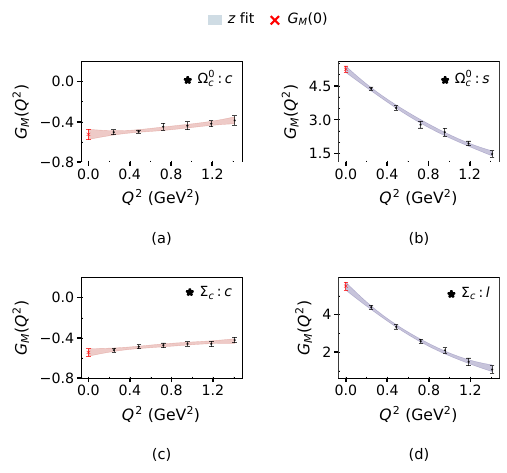}
%}
\caption{$z$-expansion fits to the flavor-separated magnetic form factors $G_M^f(Q^2)$ on H48P32. The flavor-separated form factors include all valence quarks of the indicated flavor, with the electric charge factor removed. From left to right, the upper panels show the charm- and strange-quark contributions in $\Omega_c$, and the lower panels show the charm- and light-quark contributions in $\Sigma_c$. Points are lattice results, and bands show the $z$-expansion fits and their uncertainties. The labels $c,s,l$ in the data legends identify the charm, strange, and light sectors, respectively. Red crosses show the extrapolated $G_M(0)$ values and their uncertainties.}
\label{fig:quark_M}
\end{minipage}\hfill
\begin{minipage}[t]{\columnwidth}
\centering
%\subfigure{
\includegraphics[width=\linewidth]{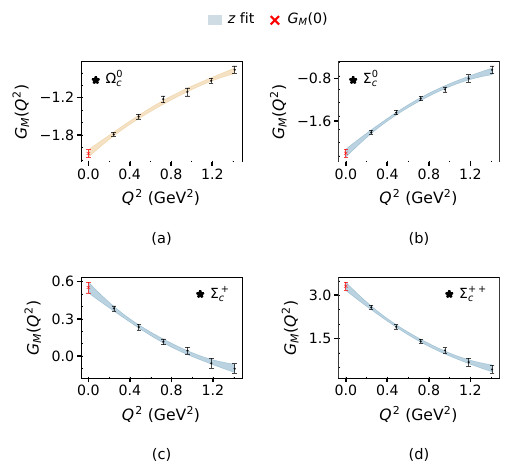}
%}
\caption{$z$-expansion fits to the total magnetic form factors on H48P32, obtained from the charge-weighted sums of the flavor contributions. From left to right, the upper panels show $\Omega_c^0$ and $\Sigma_c^0$, and the lower panels show $\Sigma_c^+$ and $\Sigma_c^{++}$. Points are lattice results, and bands show the $z$-expansion fits and their uncertainties. Red crosses show the extrapolated $G_M(0)$ values and their uncertainties.}
\label{fig:whole_M}
\end{minipage}
\end{figure*}

We fit the $Q^2$ dependence of the magnetic form factors using the same cubic $z$ expansion as for the electric form factors and extrapolate the nonzero-$Q^2$ results to determine $G_M(0)$. Figures~\ref{fig:quark_M} and~\ref{fig:whole_M} show the fits to the flavor-separated and total baryon magnetic form factors on H48P32, respectively. The fitted form factors and their slopes at $Q^2=0$ determine the flavor-separated mean-square magnetic radii through Eq.~\eqref{eq:charge_radii_z}. The flavor-separated and total baryon results are listed in Table~\ref{tab:magnetic_radii}.

The magnetic moments are determined from $G_M(0)$ and the lattice baryon masses in Table~\ref{tab:mass} using Eqs.~\eqref{eq:mu} and~\eqref{eq:total_mu}, and are expressed in nuclear magnetons.

Before multiplication by the quark electric charges, the light- and strange-quark magnetic moments are positive on all six ensembles, whereas the charm-quark magnetic moments are negative and smaller in magnitude. This sign pattern is consistent with the spin-coupling structure of the flavor sextet. In the heavy-quark limit, the angular momentum of the light degrees of freedom, $j_\ell=1$, couples to the charm-quark spin $s_c=\tfrac12$ to form total spin $J=\tfrac12$. The expectation value of the charm-quark spin is antiparallel to the baryon spin, and its magnetic coupling is suppressed by $1/m_c$. Earlier lattice studies also found that light and strange quarks dominate the magnetic moments of $\Sigma_c$ and $\Omega_c$, respectively~\cite{Can:2013tna,Can:2015exa}.

After combining the flavor contributions according to Eq.~\eqref{eq:form_factor_add}, the light- or strange-quark and charm-quark contributions in $\Sigma_c^0$ and $\Omega_c^0$ are both negative and add constructively. In $\Sigma_c^+$, the positive light-quark contribution partially cancels the negative charm-quark contribution. Our calculation gives $\mu_l>4|\mu_c|$, yielding a positive total magnetic moment. The larger electric-charge weight of the light-quark contribution in $\Sigma_c^{++}$ produces a larger positive magnetic moment.

The mean-square magnetic radius follows from the normalized slope of the total magnetic form factor at zero momentum transfer and can be written as
\begin{equation}
 \langle r_{M,\mathcal B}^2\rangle=
 \frac{\sum_f e_f G^f_{M,\mathcal B}(0)
       \langle r_{M,f}^2\rangle}
      {\sum_f e_f G^f_{M,\mathcal B}(0)}.
 \label{eq:magnetic_radius_flavor_weights}
\end{equation}
The denominator is the total magnetic form factor $G_M(0)$, and the flavor weights include the signs of both the quark charges and the magnetic form factors.

In Table~\ref{tab:magnetic_radii}, the central value of the $\Sigma_c^+$ mean-square magnetic radius is larger than those of $\Sigma_c^0$ and $\Sigma_c^{++}$ on every ensemble. This pattern arises from the partial cancellation between the light- and charm-quark contributions together with the larger mean-square magnetic radius of the light-quark contribution.

Table~\ref{tab:mu_lattice} lists the magnetic moments on each ensemble and in the continuum limit at the physical pion mass. Table~\ref{tab:mu_other} compares the extrapolated results with earlier lattice and phenomenological calculations. The combined continuum and chiral extrapolation is discussed in the next subsection.

%%%%%%%%%%%%%%%%%%%%%%%%%%%%%%%%%%%%%%%%%

%%%%%%%%%%%%%%%%%%%%%%%%%%%

%%%%%%%%%%%%%%%%%%%%%%%%%%%%%%%%%%%%

%%%%%%%%%%%%%%%%%%%%%%%%%%%%%%%%%%%%
\subsection{Continuum and chiral extrapolation}
\label{Continuum and chiral extrapolation}
We extrapolate the mean-square charge and magnetic radii and magnetic moments from the six ensembles to the continuum limit and the physical pion mass. For the baryon magnetic moments, the extrapolation uses the total moment determined on each ensemble. The flavor-separated quantities are extrapolated separately to resolve the individual quark-sector contributions. We use the fit form
\begin{eqnarray}
 f(a^2,m_\pi^2)
 &=& \langle\mathcal O\rangle(0,m_{\pi,\mathrm{phy}}^2)
     +f_1a^2 \nonumber\\
 &&+f_2(m_\pi^2-m_{\pi,\mathrm{phy}}^2).
 \label{eq:wt_}
\end{eqnarray}
Here $\mathcal O$ denotes a mean-square radius or magnetic moment, and $\langle\mathcal O\rangle(0,m_{\pi,\mathrm{phy}}^2)$ is its value in the continuum limit at the physical pion mass, $m_{\pi,\mathrm{phy}}=0.135\,\mathrm{GeV}$. The fit parameters $f_1$ and $f_2$ describe the $a^2$ and $m_\pi^2$ dependence, respectively. Equation~\eqref{eq:wt_} gives the main extrapolated results of this work. The mean-square charge radii, mean-square magnetic radii, and magnetic moments are listed in Tables~\ref{tab:electric_radii}, \ref{tab:magnetic_radii}, and~\ref{tab:mu_lattice}, respectively.

In the continuum limit at the physical pion mass, the ratios of the flavor-separated magnetic moments are $|\mu_c/\mu_s|=0.097(16)$ for $\Omega_c$ and $|\mu_c/\mu_l|=0.082(24)$ for $\Sigma_c$. These ratios quantify the suppression of the charm-quark contribution relative to the total light- and strange-quark contributions.

\begin{figure*}[!p]
\centering
\includegraphics[width=\textwidth]{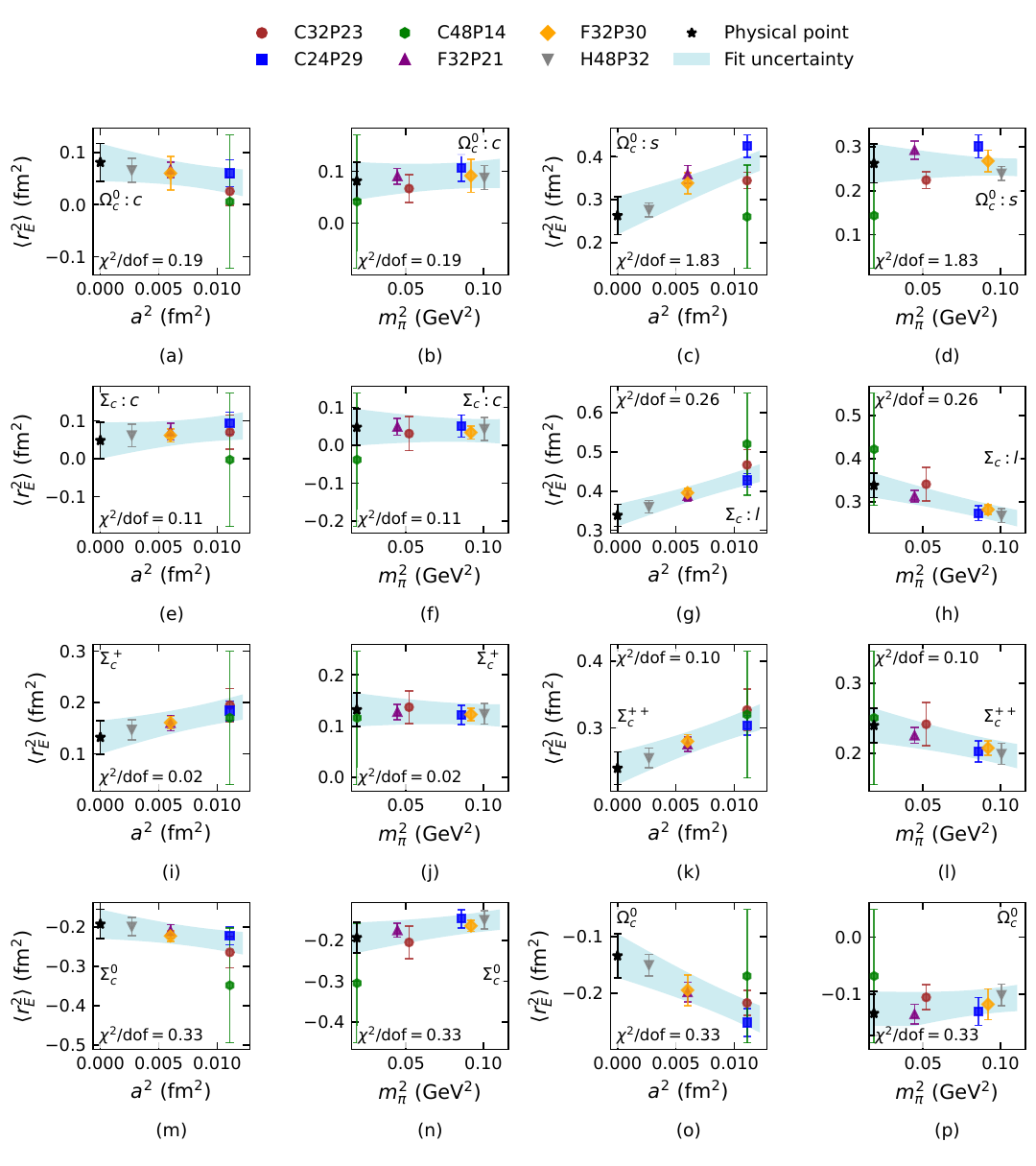}
\caption{Combined continuum and chiral extrapolations of the mean-square charge radii. Each pair of adjacent panels corresponds to one observable. From left to right, the first row shows the charm- and strange-quark contributions in $\Omega_c$, the second row shows the charm- and light-quark contributions in $\Sigma_c$, the third row shows $\Sigma_c^+$ and $\Sigma_c^{++}$, and the fourth row shows $\Sigma_c^0$ and $\Omega_c^0$. Columns 1 and 3 show the $a^2$ dependence at the physical pion mass, while columns 2 and 4 show the $m_\pi^2$ dependence in the continuum limit. In-panel labels identify the baryon and flavor. The shared legend at the top identifies the ensembles by symbol and color, black stars denote the results in the continuum limit at the physical pion mass, and bands show the fit uncertainties. For each projection, the points are shifted by subtracting the fitted contribution of the other variable.}
\label{fig:wt:ER_combined}
\end{figure*}

\begin{figure*}[!p]
\centering
\includegraphics[width=\textwidth]{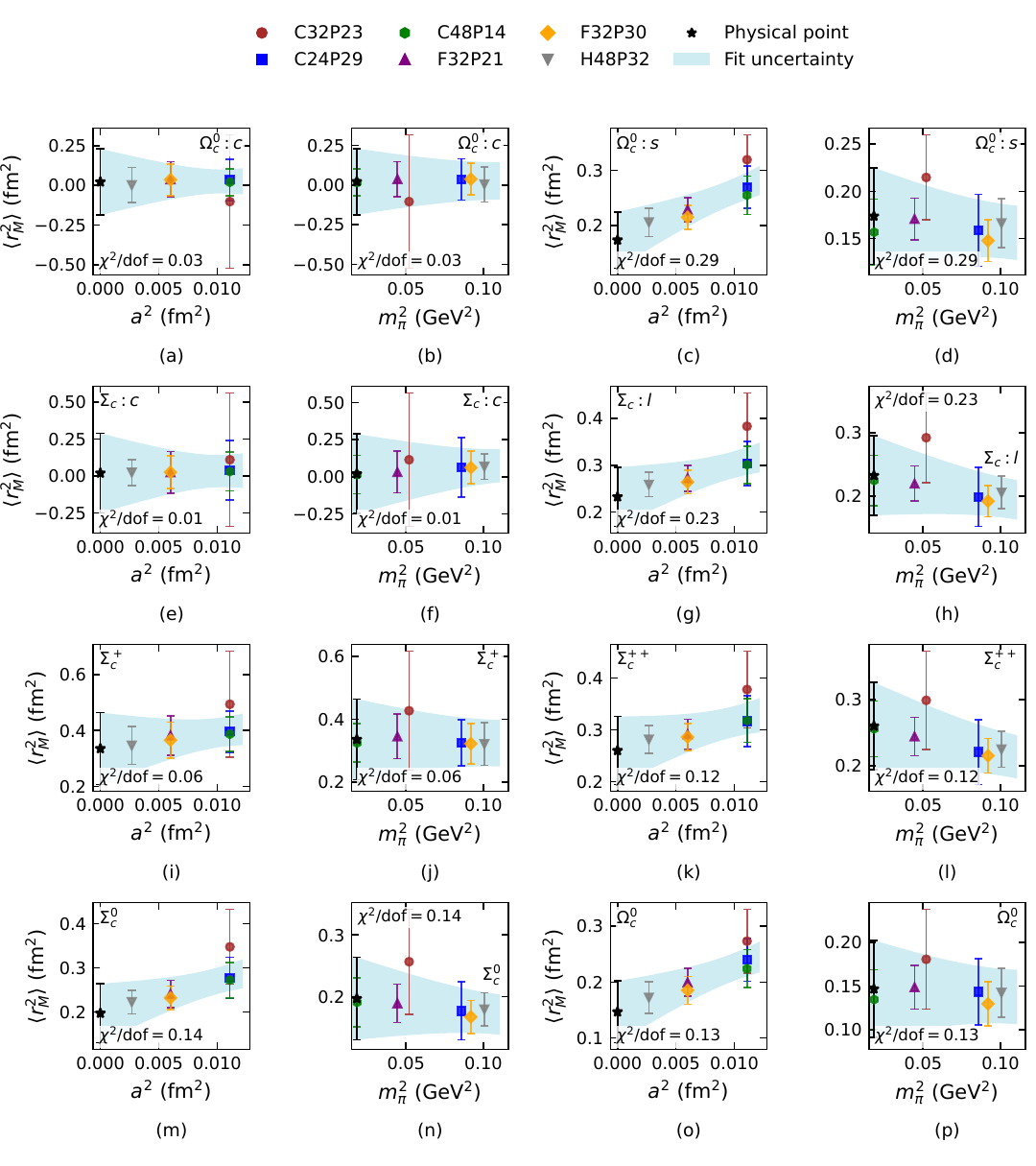}
\caption{Combined continuum and chiral extrapolations of the mean-square magnetic radii. Each pair of adjacent panels corresponds to one observable. From left to right, the first row shows the charm- and strange-quark contributions in $\Omega_c$, the second row shows the charm- and light-quark contributions in $\Sigma_c$, the third row shows $\Sigma_c^+$ and $\Sigma_c^{++}$, and the fourth row shows $\Sigma_c^0$ and $\Omega_c^0$. Columns 1 and 3 show the $a^2$ dependence at the physical pion mass, while columns 2 and 4 show the $m_\pi^2$ dependence in the continuum limit. In-panel labels identify the baryon and flavor. The shared legend at the top identifies the ensembles by symbol and color, black stars denote the results in the continuum limit at the physical pion mass, and bands show the fit uncertainties. For each projection, the points are shifted by subtracting the fitted contribution of the other variable.}
\label{fig:wt:MR_combined}
\end{figure*}

\begin{figure*}[!p]
\centering
\includegraphics[width=\textwidth]{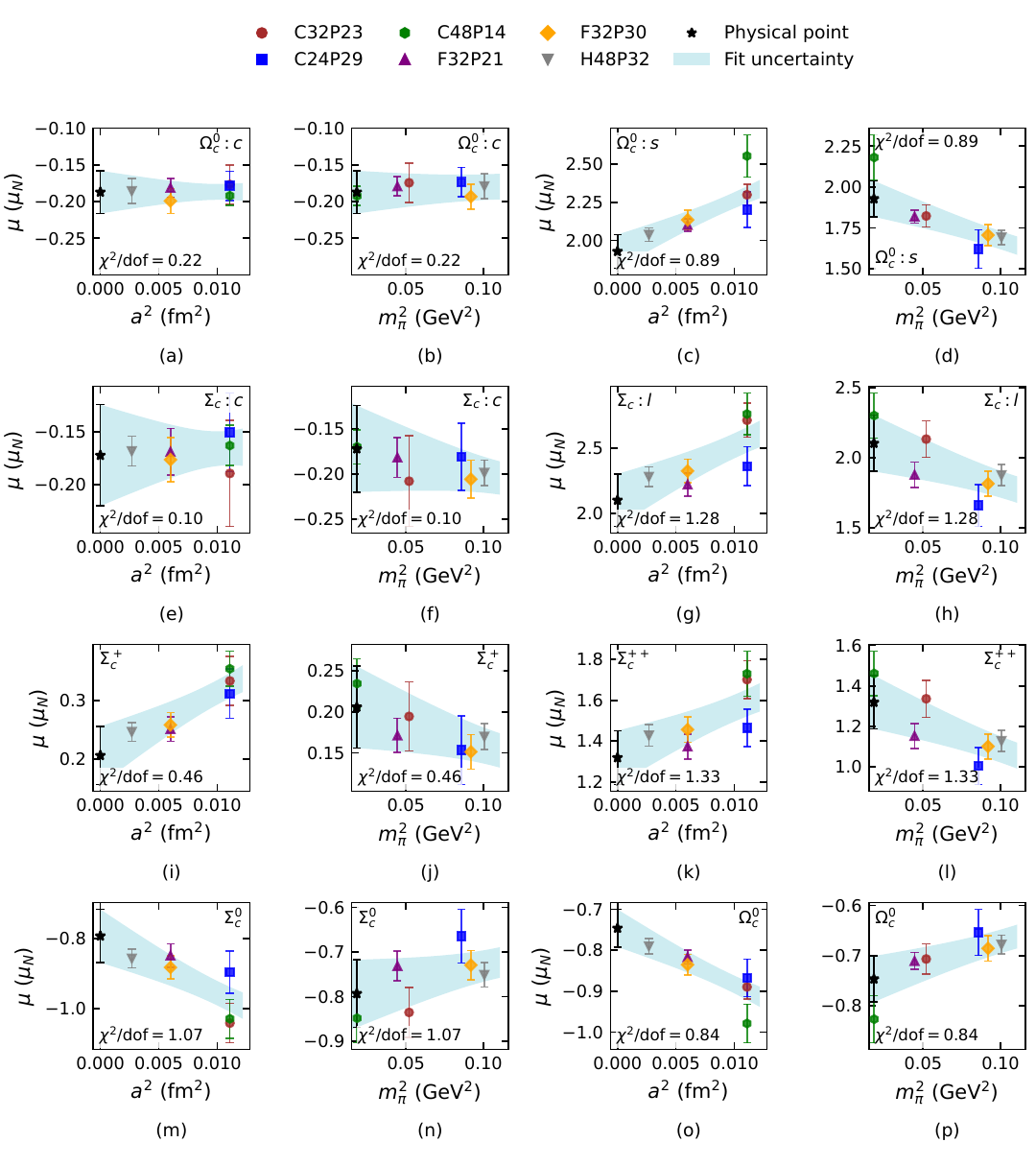}
\caption{Combined continuum and chiral extrapolations of the magnetic moments. Each pair of adjacent panels corresponds to one observable. From left to right, the first row shows the charm- and strange-quark contributions in $\Omega_c$, the second row shows the charm- and light-quark contributions in $\Sigma_c$, the third row shows $\Sigma_c^+$ and $\Sigma_c^{++}$, and the fourth row shows $\Sigma_c^0$ and $\Omega_c^0$. Columns 1 and 3 show the $a^2$ dependence at the physical pion mass, while columns 2 and 4 show the $m_\pi^2$ dependence in the continuum limit. In-panel labels identify the baryon and flavor. The shared legend at the top identifies the ensembles by symbol and color, black stars denote the results in the continuum limit at the physical pion mass, and bands show the fit uncertainties. For each projection, the points are shifted by subtracting the fitted contribution of the other variable.}
\label{fig:wt:Mmu_combined}
\end{figure*}

Figures~\ref{fig:wt:ER_combined}, \ref{fig:wt:MR_combined}, and~\ref{fig:wt:Mmu_combined} show the combined continuum and chiral extrapolations of the mean-square charge radii, mean-square magnetic radii, and magnetic moments, respectively. For each observable, the left and right panels of each pair show the lattice-spacing dependence at the physical pion mass and the pion-mass dependence in the continuum limit, respectively.

At similar pion masses, the light- and strange-quark mean-square charge radii show a clear decrease as the lattice spacing is reduced. The baryon mean-square magnetic radii show a similar downward trend. For example, the $\Omega_c^0$ mean-square magnetic radius is $0.270(57)\,\mathrm{fm}^2$ on C32P23, compared with $0.297(33)\,\mathrm{fm}^2$ in the single-spacing calculation of Can et al.~\cite{Can:2013tna} and $0.147(55)\,\mathrm{fm}^2$ from our combined continuum and chiral extrapolation. These results demonstrate the impact of the continuum extrapolation on electromagnetic radii.

%%%%%%%%%%%%%%%%%%%%%%%%%%%%%%%%%%%%%%%%%%%%%%
\begin{table*}[!p]
\centering
\caption{$\Omega_c$ and $\Sigma_c$ masses on each ensemble and their extrapolated values, compared with the PDG experimental values~\cite{ParticleDataGroup:2024cfk} and other lattice QCD results.}
\renewcommand{\arraystretch}{1.5}
\setlength{\tabcolsep}{18pt} %
\begin{tabular}{lcc}
\hline \hline
\textbf{Ensemble/source} & \textbf{$\Omega_c^0$ (GeV)} & \textbf{$\Sigma_c^{0}/\Sigma_c^{+}/\Sigma_c^{++}$ (GeV)} \\
\hline
C32P23 & 2.634(19) & 2.462(26) \\
C24P29 & 2.662(17) & 2.537(20) \\
C48P14 & 2.653(21) & 2.383(33) \\
F32P21 & 2.625(36) & 2.555(40) \\
F32P30 & 2.713(23) & 2.590(29) \\
H48P32 & 2.759(17) & 2.601(23) \\
Extrapolated& 2.733(40) & 2.486(50) \\
PDG~\cite{ParticleDataGroup:2024cfk} & 2.6953(4) & 2.45375(14) \\
\hline
\citet{Can:2013tna} & 2.740(24) & 2.553(18) \\
\citet{Hu:2024mas} & 2.6985(123)(38) & 2.4574(201)(197) \\
\citet{PACS-CS:2013vie} & 2.673(13) & 2.467(40) \\
\citet{Brown:2014ena} & 2.679(37)(20) & 2.474(41)(25) \\
\citet{Briceno:2012wt} & 2.681(31)(12)(5) & 2.481(24)(15)(7) \\
\hline \hline
\end{tabular}
\label{tab:mass}
\end{table*}

\begin{table*}[!p]
\centering
\caption{Mean-square charge radii of singly charmed baryons on the six ensembles and in the continuum limit at the physical pion mass, compared with earlier lattice QCD results. All values are in $\mathrm{fm}^2$. For neutral baryons, the mean-square charge radius is obtained from the slope at $Q^2=0$ of the total electric form factor constructed using Eq.~\eqref{eq:form_factor_add}.}
\label{tab:electric_radii}
\renewcommand{\arraystretch}{1.4}
\small
\begin{tabular*}{\textwidth}{@{\extracolsep{\fill}}lccccccccc@{}}
\hline \hline
& \multicolumn{3}{c}{$\Omega_c^0$} & \multicolumn{5}{c}{{$\Sigma_c^{0}/\Sigma_c^{+}/\Sigma_c^{++}$}} \\
\hline
Ensemble & $r^2_{E,c}$ & $r^2_{E,s}$ & $r^2_{E,\Omega_c^0}$ & $r^2_{E,c}$ & $r^2_{E,l}$ & $r^2_{E,\Sigma_c^{0}}$ & $r^2_{E,\Sigma_c^{+}}$ & $r^2_{E,\Sigma_c^{++}}$ \\
\hline
C32P23 & 0.031(27) & 0.341(19) & -0.207(22) & 0.066(45) & 0.439(39) & -0.249(40) & 0.191(32) & 0.311(31) \\
C24P29 & 0.071(26) & 0.418(26) & -0.231(25) & 0.086(29) & 0.372(17) & -0.191(22) & 0.176(19) & 0.272(15) \\
C48P14 & 0.006(128) & 0.26(12) & -0.17(12) & -0.003(176) & 0.52(13) & -0.35(15) & 0.17(13) & 0.320(95)  \\
F32P21 & 0.071(15) & 0.356(21) & -0.190(17) & 0.068(22) & 0.366(13) & -0.199(17) & 0.157(15) & 0.263(11) \\
F32P30 & 0.072(32) & 0.331(25) & -0.173(27) & 0.053(17) & 0.335(11) & -0.188(13) & 0.152(12) & 0.245(10) \\
H48P32 & 0.079(23) & 0.268(16) & -0.126(19) & 0.052(30) & 0.292(16) & -0.160(23) & 0.137(20) & 0.216(15) \\
\hline
Extrapolated & 0.081(36) & 0.263(44) & -0.134(38) & 0.048(48) & 0.338(28) & -0.193(37) & 0.132(32) & 0.240(24) \\
\citet{Can:2013tna} & 0.064(10) & 0.330(32) & - & 0.032(18) & 0.347(49) & - & - & 0.192(22) \\
\citet{Can:2015exa} & 0.064(11) & 0.329(25) & -0.177(18) & - & - & - & - & - \\
\hline \hline
\end{tabular*}
\end{table*}

\begin{table*}[!p]
\centering
\caption{Mean-square magnetic radii of singly charmed baryons on the six ensembles and in the continuum limit at the physical pion mass, compared with earlier lattice QCD results. All values are in $\mathrm{fm}^2$.}
\setlength{\tabcolsep}{0.2cm}
\renewcommand{\arraystretch}{1.2}
\small
\resizebox{\textwidth}{!}{%
\begin{tabular}{@{\extracolsep{0.8em}} lcccccccc }
\hline \hline
 & \multicolumn{3}{c}{$\Omega_c^0$} & \multicolumn{5}{c}{$\Sigma_c^{0}/\Sigma_c^{+}/\Sigma_c^{++}$} \\ \hline
Ensemble & $r^2_{M,c}$ & $r^2_{M,s}$ & $r^2_{M,\Omega_c^0}$ & $r^2_{M,c}$ & $r^2_{M,l}$ & $r^2_{M,\Sigma_c^0}$ & $r^2_{M,\Sigma_c^{+}}$ & $r^2_{M,\Sigma_c^{++}}$ \\
\hline
C32P23 & -0.10(42) & 0.313(45) & 0.270(57) & 0.13(45) & 0.369(71) & 0.338(85) & 0.49(19) & 0.361(74) \\
C24P29 & 0.04(13) & 0.257(38) & 0.233(38) & 0.08(20) & 0.275(47) & 0.258(47) & 0.388(74) & 0.283(49) \\
C48P14 &0.020(85) & 0.255(35) & 0.224(34)&0.03(13)&0.301(40)&0.272(40)&0.387(61)&0.318(42)\\
F32P21 & 0.04(11) & 0.224(22) & 0.197(25) & 0.04(14) & 0.261(28) & 0.233(31) & 0.379(71) & 0.278(29) \\
F32P30 & 0.04(10) & 0.201(22) & 0.178(25) & 0.07(11) & 0.233(25) & 0.211(27) & 0.356(65) & 0.249(26) \\
H48P32 & 0.006(111) & 0.190(26) & 0.164(28) & 0.073(87) & 0.224(26) & 0.199(27) & 0.336(68) & 0.240(27) \\ \hline
Extrapolated & 0.02(21) & 0.174(51) & 0.147(55) & 0.02(27) & 0.232(63) & 0.197(67) & 0.34(13) & 0.260(66)\\
\citet{Can:2013tna} & 0.056(19) & 0.398(44) & 0.297(33) & 0.098(80) & 0.403(67) & 0.377(75) & - & 0.410(81) \\
\hline \hline
\end{tabular}}
\label{tab:magnetic_radii}
\end{table*}

\begin{table*}[!p]
\centering
\caption{Magnetic moments of singly charmed baryons in nuclear magnetons, on the six ensembles and in the continuum limit at the physical pion mass, compared with earlier lattice QCD results. The flavor-separated moments include the full contribution of each flavor current before multiplication by the quark electric charge. The full flavor contributions quoted in Ref.~\cite{Can:2013tna} are used as published. The strange-quark magnetic moment and its uncertainty from Ref.~\cite{Can:2015exa} are multiplied by two to match the normalization used here.}
\setlength{\tabcolsep}{0.2cm}
\renewcommand{\arraystretch}{1.2}
\small
\resizebox{\textwidth}{!}{%
\begin{tabular}{@{\extracolsep{0.8em}} lcccccccc }
\hline \hline
 & \multicolumn{3}{c}{$\Omega_c^0$} & \multicolumn{5}{c}{$\Sigma_c^{0}/\Sigma_c^{+}/\Sigma_c^{++}$} \\ \hline
Ensemble & $\mu_{c}$ & $\mu_{s}$ & $\mu_{\Omega_c^0}$ & $\mu_{c}$ & $\mu_{l}$ & $\mu_{\Sigma_c^0}$ & $\mu_{\Sigma_c^{+}}$ & $\mu_{\Sigma_c^{++}}$ \\
\hline
C32P23 & -0.174(27) & 2.194(68) & -0.859(30) & -0.201(50) & 2.59(13) & -1.016(56) & 0.314(42) & 1.605(92) \\
C24P29 & -0.173(20) & 1.99(12) & -0.806(46) & -0.174(37) & 2.12(15) & -0.845(60) & 0.273(42) & 1.273(92) \\
C48P14 & -0.192(13) & 2.55(14) & -0.979(47) & -0.163(19) & 2.76(16) &-1.029(55) &0.354(30)& 1.73(11) \\
F32P21 & -0.179(13) & 2.020(40) & -0.793(17) & -0.178(22) & 2.128(90) & -0.829(33) & 0.236(21) & 1.298(61) \\
F32P30 & -0.193(17) & 1.906(64) & -0.768(25) & -0.202(21) & 2.064(90) & -0.827(33) & 0.216(21) & 1.246(62) \\
H48P32 & -0.179(17) & 1.782(44) & -0.715(19) & -0.197(14) & 1.988(76) & -0.795(27) & 0.199(16) & 1.193(52) \\
\hline
Extrapolated & -0.187(29) & 1.93(11) & -0.746(46) & -0.172(48) & 2.10(20) & -0.793(76) & 0.206(50) & 1.32(13) \\
\hline
\citet{Can:2013tna} & -0.099(14) & 1.71(15) & -0.608(45) & -0.099(21) & 2.37(36) & -0.852(133) & - & 1.569(253) \\
\citet{Can:2015exa} & -0.092(6) & 1.958(94) & -0.688(31) & - & - & - & - & - \\
\hline \hline
\end{tabular}}
\label{tab:mu_lattice}
\end{table*}

\begin{table*}[!p]
\centering
\caption{Comparison of magnetic moments with other theoretical calculations, in nuclear magnetons. EOMS and HB denote the extended on-mass-shell renormalization scheme and heavy-baryon chiral perturbation theory, respectively.}
\label{tab:mu_other}
\renewcommand{\arraystretch}{1.2}
\small
\begin{tabular*}{\textwidth}{@{\extracolsep{\fill}}lcccc@{}}  %
\hline \hline
\textbf{} & \textbf{$\mu_{\Omega_c^0}$} & \textbf{$\mu_{\Sigma_c^0}$} & \textbf{$\mu_{\Sigma_c^+}$} & \textbf{$\mu_{\Sigma_c^{++}}$} \\
\hline
This work & $-0.746(46)$ & $-0.793(76)$ & $0.206(50)$ & $1.32(13)$ \\
\hline
\citet{Can:2013tna} & $-0.608(45)$ & $-0.852(133)$ & - & $1.569(253)$ \\
\citet{Can:2015exa} & $-0.688(31)$ & - & - & - \\
\hline
\citet{Mohan:2022sxm} & $-0.905$ & $-1.234$ & $0.432$ & $2.095$ \\
\citet{Simonis:2018rld} & $-0.950$ & $-1.31$ & $0.487$ & $2.28$ \\
\citet{Wang:2018gpl} & $-0.69$ & $-0.97$ & $0.26$ & $1.50$ \\
\citet{Shi:2018rhk} EOMS & $-0.69$ & $-0.83$ & $0.27$ & $1.38$ \\
\citet{Shi:2018rhk} HB & $-0.74$ & $-1.08$ & $0.46$ & $2.00$ \\
\citet{Aliev:2015axa} & $-0.9$ & $-1.5$ & $0.5$ & $2.4$ \\
\citet{Yang:2018uoj} & $-0.85(5)$ & $-1.24(5)$ & $0.46(3)$ & $2.15(10)$ \\
\citet{Faessler:2006ft} & $-0.85$ & $-1.04$ & $0.36$ & $1.76$ \\
\citet{Zhu:1997as} & $-$ & $-1.6(2)$ & $0.23(3)$ & $2.1(3)$ \\
\citet{Kucukyilmaz:2025rsd} & $-0.949$ & $-$ & $-$ & $-$ \\
\citet{Sharma:2010vv} & $-0.90$ & $-1.60$ & $0.30$ & $2.20$ \\
\hline \hline
\end{tabular*}

\end{table*}

%%%%%%%%%%%%%%%%%%%%%%%%%%%%%%%%%%%%%%%%%%%%%%

%%%%%%%%%%%%%%%%%%%%%%%%%%%%%%%%%%%%%%%%%%%%%%%%%%%
%%%%%%%%%%%%%%%%%%%%%%%%%%%%%%

\clearpage
\section{Conclusions}
\label{sec:conclusions}
We have calculated the electric and magnetic form factors of $\Sigma_c^0$, $\Sigma_c^+$, $\Sigma_c^{++}$, and $\Omega_c^0$ on six lattice ensembles. Using the $z$ expansion and combined continuum and chiral extrapolations, we have determined their mean-square charge and magnetic radii, magnetic moments, and flavor contributions in the continuum limit at the physical pion mass.

The flavor-separated mean-square charge radii show that the light- and strange-quark charge distributions are more extended than the charm-quark distribution, consistent with a compact heavy-quark color source surrounded by more extended light degrees of freedom. In $\Sigma_c^0$ and $\Omega_c^0$, the more extended distributions of the negatively charged light or strange quarks give rise to negative mean-square charge radii.

The flavor-separated magnetic moments are positive for the light and strange sectors and negative, with smaller magnitudes, for the charm sector. In the continuum limit at the physical pion mass, the ratios are $|\mu_c/\mu_s|=0.097(16)$ for $\Omega_c$ and $|\mu_c/\mu_l|=0.082(24)$ for $\Sigma_c$. The $\Sigma_c^0$ and $\Omega_c^0$ magnetic moments are negative. Partial cancellation between the light- and charm-quark contributions gives a small positive magnetic moment for $\Sigma_c^+$, whereas $\Sigma_c^{++}$ has a larger positive magnetic moment.

The light- and strange-quark mean-square charge radii exhibit a clear lattice-spacing dependence. The baryon mean-square magnetic radii also show a decreasing trend as the lattice spacing is reduced. The combined continuum and chiral extrapolation yields smaller mean-square magnetic radii than the coarse-lattice results. The continuum results and flavor contributions provide quantitative input for testing theoretical descriptions of the electromagnetic structure of singly charmed baryons.

\begin{acknowledgments}
This work is supported by the National Natural Science Foundation of China under Grants No.~12675103, No.~12605147, No.~12575085, and No.~12635009. W.~Qin is supported by the Natural Science Foundation of Hunan Province under Grant No.~2024JJ6300 and the Hunan Provincial Department of Education under Grant No.~22B0044. L.~Lu is supported by the Shandong Provincial Natural Science Foundation under Grant No.~ZR2023QA132. We acknowledge the use of Chroma~\cite{Edwards:2004sx}, QUDA~\cite{Clark:2009wm,Babich:2011np}, and PyQUDA~\cite{Jiang:2024lto}. Numerical calculations were performed on the Xiangjiang-1 cluster at Hunan Normal University in Changsha, the Southern Nuclear Science Computing Center (SNSC), the high-performance computing cluster at the Institute of High Energy Physics in Beijing, the high-performance computing cluster at the China Spallation Neutron Source in Dongguan, and the ORISE supercomputer.

\end{acknowledgments}

\interlinepenalty=10000
\bibliography{my_references}

\end{document}